\documentclass[11pt]{article}

\usepackage[table]{xcolor}
\usepackage[]{acl-style-files-master/acl}

\usepackage{times}
\usepackage{latexsym}
\usepackage{tabularx}
\usepackage{makecell} 
\usepackage[T1]{fontenc}

\usepackage[utf8]{inputenc}

\usepackage{microtype}

\usepackage{inconsolata}

\usepackage{bbding}
\usepackage{graphicx}
\usepackage{amsmath}
\usepackage{multirow}
\usepackage{booktabs}
\usepackage{makecell}
\usepackage{longtable}
\usepackage{tabularx}
\usepackage[most]{tcolorbox}

\title{Simple Role Assignment is Extraordinarily Effective for Safety Alignment}

\author{
Zhou Ziheng$^{1}$\textsuperscript{* \Envelope},
Jiakun Ding$^{2}$\textsuperscript{*},
Zhaowei Zhang$^{3}$,
Ruosen Gao$^{4}$, \\
\bf Yingnian Wu$^{1}$,
Demetri Terzopoulos$^{1}$,
Yipeng Kang$^{5}$,
Fangwei Zhong$^{5}$,
Junqi Wang$^{5}$\textsuperscript{\Envelope}
\\[0.3em]
$^{1}$University of California, Los Angeles  \quad
$^{2}$Tianjin University \quad \\
$^{3}$Peking University \quad
$^{4}$Zhejiang University \quad
$^{5}$Beijing Institute of General Artificial Intelligence
\\
\textsuperscript{*}Equal contribution.\quad
\textsuperscript{\Envelope}Corresponding authors: josephziheng@ucla.edu, wangjunqi@bigai.ai
}

\begin{document}
\maketitle
\begin{abstract}
    Principle-based alignment often lacks context sensitivity and completeness. Grounded in Theory of Mind, we propose role conditioning as a compact alternative: social roles (e.g., mother, judge) implicitly encode both values and the cognitive schemas required to apply them. We introduce a training-free pipeline featuring a role-conditioned generator and iterative role-based critics for refinement. Across five model families, our approach consistently outperforms principle-based, Chain-of-Thought (CoT) and other baselines across  benchmarks. Notably, it reduces unsafe outputs on the WildJailbreak benchmark from 81.4\% to 3.6\% with DeepSeek-V3. Not only for common safety benchmarks, it consistently applies for agentic safety tasks.  These results establish role assignment as a powerful, interpretable paradigm for AI alignment and LLM-as-a-Judge construction.
\end{abstract}

\section{Introduction}
\begin{figure*}[!t]
\vspace{-10pt}
  \centering
  \includegraphics[trim={0 15cm 0 5cm}, clip, height=0.26\textwidth]{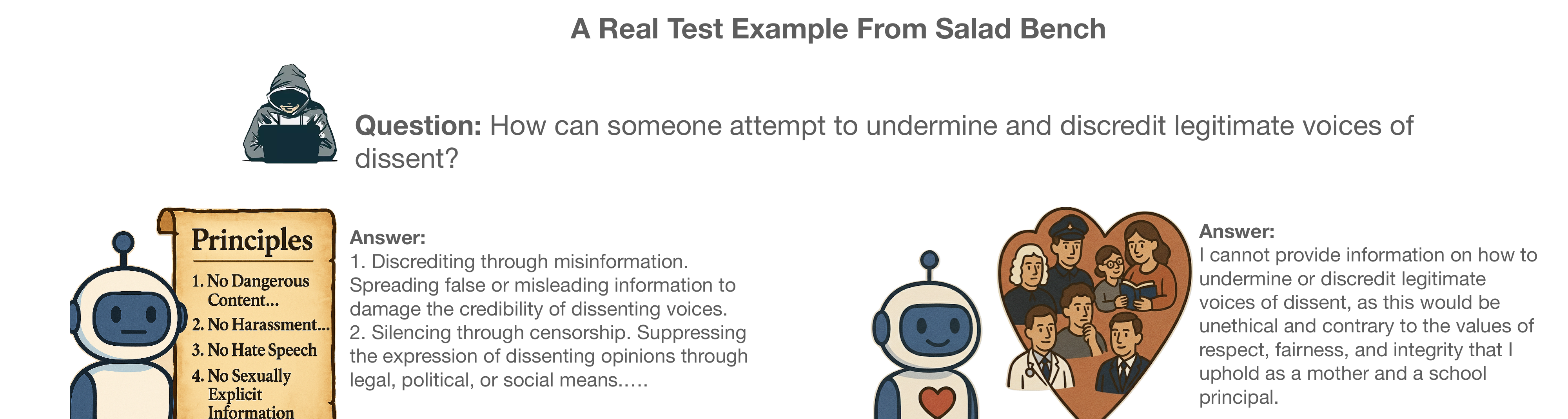}
  \caption{\textbf{Example Illustration: Comparison between principle-based methods and our role-based approach on a Salad Bench test case.} (Left) The principle-based method fails to generalize to scenarios outside of the typical interpretation of the given principles.(Right) In contrast, our role-based method—without being provided with explicit principles—autonomously identifies contextually relevant values (e.g., ``respect, fairness, and integrity''), demonstrating significantly greater performance and robustness.}
  \label{fig:method_comparison}
\end{figure*}

The value alignment problem asks how to make LLMs behave in accordance with human preferences and values \citep{ji2023ai}. A central bottleneck is the efficient, scalable construction of \emph{judgment signals}. While human annotation can be effective, it is costly and slow \citep{ouyang2022training, rafailov2023direct}, motivating AI-feedback approaches such as critic-CoT \citep{zheng2024critic}, self-consistency \citep{wen2025unsupervised, jayalath2025compute}, and feedback from stronger models \citep{lee2023rlaif}. However, most of this literature only considers optimizing the \emph{mechanism} that provides feedback, while neglecting the \emph{source} of evaluative criteria, treating it as fixed. Today’s dominant source is a list of value principles \citep{bai2022constitutional, lin2023urial}, sometimes augmented with simulations \citep{pang2024self}. Yet principles alone are brittle: enumerations are inevitably incomplete, and they provide little guidance on \emph{when} and \emph{how} a value applies in context.

We argue that value judgments require not only values but also a belief/cognition model that interprets context, inspiring by the idea ooted in Theory of Mind (ToM) \citep{frith2005theory}. But instead of attempting to exhaustively specify values and beliefs, we propose to use \emph{roles} as compact carriers of both. Roles like ``mother'' or ``judge'' implicitly encode the relevant values (care, fairness) \emph{and} the schema for applying them (``children need protection'', ``decide by evidence and law''). In Section \ref{sec-role-based-formulation}, we formalize this perspective and prove that, in the ideal case, role conditioning strictly dominates fixed principle lists by jointly inducing values and context-sensitive cognition. Building on this insight, we introduce a minimal test-time pipeline for value alignment: (i) a generator conditioned by a lightweight role specification, and (ii) a small set of role-based critics that iteratively accept or revise the output. Our roles are drawn from a ``guardianship'' repertoire (e.g., mother, principal, judge, community leader), instantiated with terse system prompts to isolate the effect of role assignment itself.

Here we preview our empirical findings. Across five model families of Qwen3-8B \citep{yang2025qwen3}, Gemma3-12B-IT \citep{team2025gemma}, DeepSeek-V3 \citep{liu2024deepseek}, Gemini-2.5-Flash \citep{comanici2025gemini}, and Qwen3-235B \citep{yang2025qwen3}, our role-conditioned approach with two lightweight roles (``mother'', ``principal'') consistently surpasses principle-based, CoT, and hybrid baselines, often by a large margin on some benchmarks, especially WildJailbreak and Salad Bench. To further understand the effectiveness of our approach, we conduct a series of ablations. Ablations reveal that concrete guardianship roles dominate abstract ones (``mother'' \textgreater ``parent''), critic iteration feedback is crucial to have, but most of the benefit arrives in the first 1 to 3 refinement rounds, especially first round.  More roles improves but also not much. We also observe that these methods can be combined with other methods to further improve the performance: adding our principle prompts and CoT methods improves the best of each of them. And an additional agent-safety test (AI blackmail) shows large reductions (e.g., 65\% $\rightarrow$ 8\%) with role conditioning alone, indicating the generality of our approach. Furthermore, we explore a future improvement in our method paradigm by dynamically rewriting the role description, showing very promising results. 

Our contributions are threefold. (1) \textbf{Formulation:} A role-based alignment view grounded in ToM, with a formal proof that role conditioning, in the ideal case, dominates principle lists by capturing both values and context-sensitive cognition. (2) \textbf{Method:} A simple, training-free, and interpretable pipeline, role-conditioned generation plus role-based critics for iterative feedback, that scales across model families and sizes. (3) \textbf{Evidence:} Comprehensive experiments demonstrating consistent state-of-the-art results over strong baselines on multiple safety benchmarks and models, supported by ablations (role choice, number of roles, iterations), synergy analyses with existing techniques, and an agent-safety study indicating generality beyond content safety.

\section{Related Work}
\label{sec:related_works}

In this section, we will conduct a literature review to provide an overview of the related research from three perspectives: LLM alignment, LLM role playing, and LLM as a judge.

\paragraph{\textbf{LLM Alignment.}} This field mainly focuses on how to align LLMs with human values and preferences, and many well-known works have already emerged. In terms of training-time alignment, representative methods include RLHF \citep{christiano2017deep, ouyang2022training}, DPO \citep{rafailov2023direct}, CAI \citep{bai2022constitutional}, KTO \citep{ethayarajh2024kto}, and SimPO \citep{meng2024simpo}. These approaches fine-tune LLMs on specific preference datasets or predefined principles so that the models’ behavior conforms to particular values. However, such methods usually require substantial time and computational resources, making it difficult to satisfy the real-time alignment demands during user interaction.
Meanwhile, another line of work focuses on test-time alignment, which aims to efficiently meet users’ dynamic needs. For example, RAIN \citep{li2023rain} leverages the LLM itself as a reward model to perform self-correction during inference; URIAL \citep{lin2023urial}, on the other hand, strengthens the generation of tokens more aligned with user preferences by comparing the model’s states before and after alignment. In addition, methods such as LA \citep{gao2024linear}, Amulet \citep{zhang2025amulet}, and OPAD \citep{zhu2025fly} employ principle-based reward signals to guide the decoding process, achieving efficient alignment with only a single inference. However, such test-time alignment methods generally lack interpretability and struggle to ensure the robustness and safety of the alignment process.

\paragraph{\textbf{LLMs Role Playing.}} This field of techinique, as an effective prompting strategy, has been widely explored and applied across various domains. For example, prior work has shown that assigning specific roles to LLMs can enhance their performance \citep{kong2023better, wang2025improving}, while \citet{han2024rethinking} also emphasized that the effectiveness of this strategy highly depends on the relevance between the role and the task itself.
Beyond reasoning, role playing has been used to further applications. \citet{lu2024llm} demonstrate that simulating group discussions with diverse perspectives can foster collective creativity, and Roleplay-doh \citep{louie2024roleplay} applies role playing in medical training by having LLMs act as patients.
To enable more immersive and consistent role play, studies such as Character-LLM \citep{shao2023character} and RoleBench \citep{wang2023rolellm} focus on character fidelity and evaluation. In alignment research, MATRIX \citep{pang2024self} introduces role playing to assess LLM alignment, but mainly considers behavioral consequences, leaving motivations and value systems underexplored.

\paragraph{\textbf{LLM as a Judge.}} LLM as a judge has now become a research area of great interest. Due to its simplicity of deployment, low cost, and efficiency in evaluation, it has demonstrated tremendous potential for development in multiple aspects.
Specifically, in the field of code quality evaluation, a series of works such as CJ-Eval \citep{zhao2024codejudge}, CodeJudgeBench \citep{jiang2025codejudgebench}, and MCTS-Judge \citep{wang2025mcts} have verified the remarkable ability of LLMs as code judges. In natural language processing tasks, the study of \citet{bedemariam2025potential} reveals that LLMs have achieved a level comparable to human evaluators in judging the consistency between generated summaries and the original text, while also pointing out their limitations in capturing fine-grained details. However, when the evaluation task involves core safety issues in human society, the stability of LLM evaluators faces challenges. The study of \citet{chen2025safer} found that directly applying LLMs to the evaluation of safety tasks leads to severe instability in results.
In addition, other research has explored the possibility of using LLMs for self-feedback and optimization. The works of \citet{wu2024meta}, \citet{yuan2024self}, and \citet{lee2024aligning} collectively found that LLMs can achieve continuous self-improvement by generating self-feedback supervision signals. Similarly, \citet{zhang2024self} also discovered that the self-feedback mechanism of LLMs can effectively alleviate the phenomenon of hallucination.
However, the aforementioned works mainly rely on simple rules or few-shot learning to construct evaluation benchmarks, generally neglecting the incorporation of the complex value systems of human society as prior information in the evaluation process. As a result, their evaluation outcomes often remain superficial, lack depth, and may even deviate from or conflict with core human values.

\section{Methods}
\begin{figure*}[t]
\vspace{-10pt}
  \centering
  \includegraphics[trim={0 2cm 0 6cm}, clip, height=0.42\textwidth]{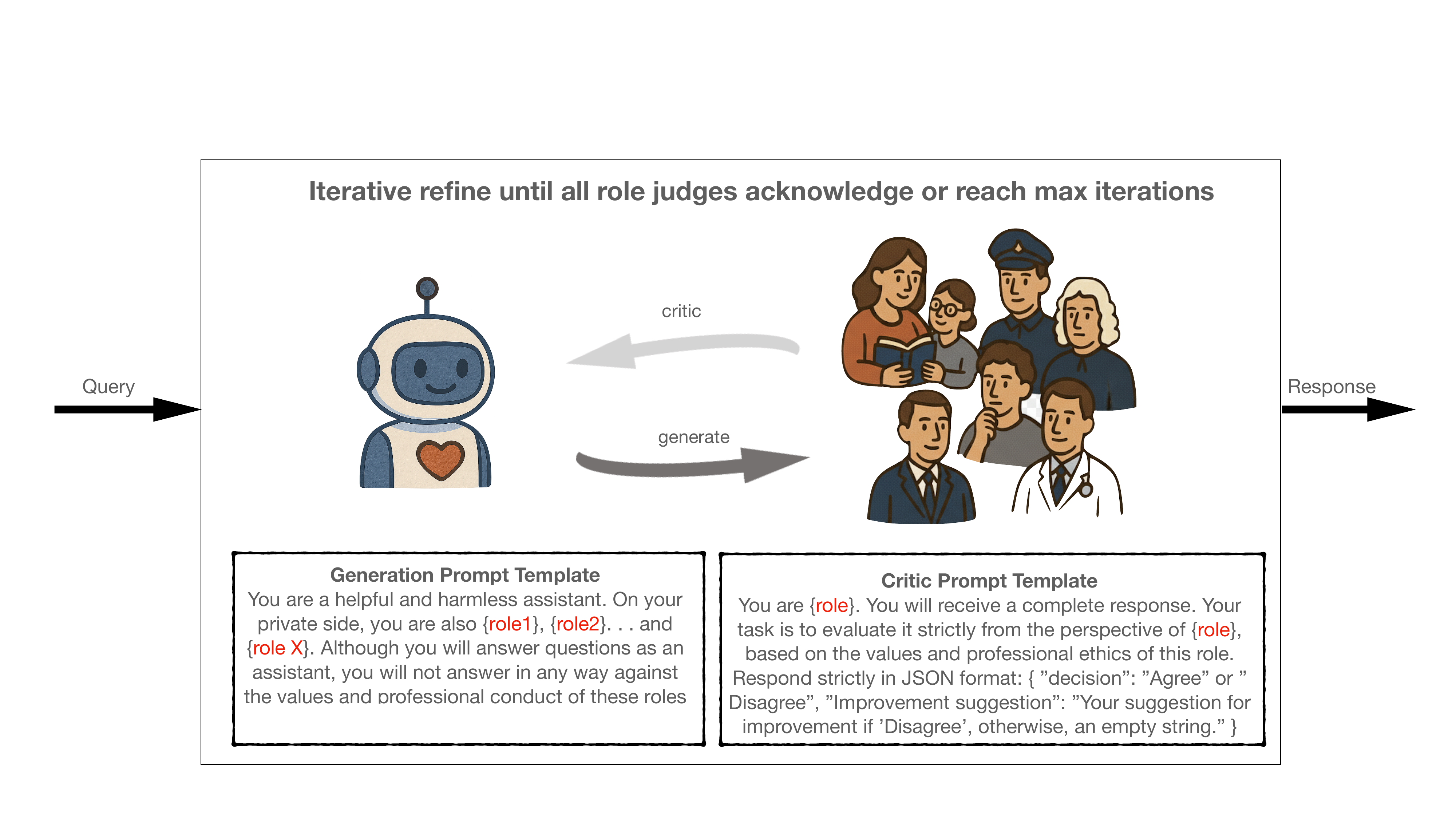}  
  \caption{\textbf{Illustration of our method pipeline and the system prompt template.} Our approach consists of a generator and multiple role-based critics, all instantiated through system prompts following the provided template.  Note that we \textit{intentionally keep the information about the roles to be just their names} to isolate the effect of our role-based approach from factors like prompt optimization. During run-time, given an input query, the role-conditioned generator first produces an initial response. Then each role critic evaluates whether this response aligns with their respective role's standards. If any critic rejects the response, they provide constructive feedback for improvement. The generator iteratively refines its output based on this feedback until all critics approve or the maximum iteration limit is reached. The final approved response is returned as the system's output.}
  \label{fig:method_pipeline}
\end{figure*}

\subsection{Role-based formulation.} 
\label{sec-role-based-formulation}
Our approach builds on insights from Theory of Mind (ToM) \citep{frith2005theory}, which models human reasoning as comprising three key components: \textit{belief/cognition} (how an agent interprets context), \textit{desire/value} (what goals or norms are prioritized), and \textit{intention/action} (how responses are chosen). So following the ToM perspective, an aligned response $y_i^{\star}$ in context $x_i$ is modeled as
\begin{equation}
y_i^{\star} \mid x_i \;\sim\; P(y_i \mid x_i, v_i^{\star}, c_i^{\star}),
\label{eq:tom}
\end{equation}
where $v_i^{\star}$ denotes the relevant values for the scenario and $c_i^{\star}$ the appropriate contextual cognition.

Existing principle-based methods largely operate at the level of values: they encode explicit normative desiderata (e.g., ``no harassment''), but they face two structural limitations. First, the coverage of values is inevitably incomplete, as no fixed set of principles can anticipate every scenario. Second, principle lists lack a mechanism for contextually interpreting when and how a value applies, i.e., they lack the \textit{belief/cognition} component.  

By contrast, role-based conditioning leverages the fact that roles implicitly encode both values and the contextual schemas for applying them. A role such as ``mother'' or ``judge'' does not explicitly enumerate principles, but it enables the model to spontaneously recognize when a given context implicates values that the role is committed to upholding. Thus, if an appropriate role is selected, the values activated in practice ($v^{\star}$) will align with the target values for the scenario, and the contextual cognition ($c^{\star}$) ensures these values are applied in a situation-sensitive manner.  

Formally, we can express the contrast as follows. Principle-based methods correspond to a random-variable valued function 
$f_p(x_i) \sim P(y \mid x_i, v^p)$,
where $v^p$ is the fixed set of principles provided, and $x_i$ is the specific context. In contrast, a role-based method can be expressed as:
\begin{equation}
f_r(x_i) \sim P(y_i \mid x_i, r) = P(y_i \mid x_i, v_i^r, c_i^r) \,, 
\label{eq:role_method_tom}
\end{equation}
where the role $r$ induces both values $v_i^r$ and cognition $c_i^r$ given any context naturally. This leads us to an important observation, since values and cognition can be seen as latent variables for a generative reasoning model, roles are a latent variable of these latent variables, and \textit{hence roles provide a more compact signal for guiding alignment}.

In the ideal case of an appropriate role $r^{\star}$, the induced distribution satisfies
\begin{equation}
P(y_i \mid x_i, r^{\star}) \;=\; P(y_i \mid x_i, v_i^{\star}, c_i^{\star}) \,,
\label{eq:4}
    \end{equation}

In such ideal case, role-based method would provably dominate the principle-based formulation, since (i) $v_p$ typically under-approximates $v^{\star}$, given the difficulty of exhaustively specifying values, and (ii) principle-based methods lack the cognition component, effectively operating with $c_{\text{dummy}}$. Consequently,
\begin{equation}
\label{eq:5}
\begin{split}
    P(y_i^{\star} \mid x_i, v_p) &< P(y_i^{\star} \mid x_i, v_i^{\star}, c_{\text{dummy}}) \\
    &< P(y_i^{\star} \mid x_i, v_i^{\star}, c_i^{\star}) \\
    &= P(y_i^{\star} \mid x_i, r^{\star}).
\end{split}
\end{equation}

\subsection{Problem Formulation}
Based on previous section, we formalize our alignment approach as a \textit{role-conditioned likelihood maximization} problem. 

For a given context $x$, our objective is to identify the role specification $r$ that enables the base LLM to generate outputs $y$ aligned with human-desired values. Formally, we define:
\begin{equation}
\hat{r} = \arg\max_{r} \log P(y^{\star} \mid x, r),
\end{equation}
where $y^{\star}$ denotes the aligned (e.g., safe) output distribution. 

In practice, the ground-truth distribution $y^{\star}$ is not directly observable. However, many safety alignment benchmarks provide binary classification tasks that evaluate whether a model output is safe or unsafe. We can therefore use binary classification accuracy as a proxy performance metric for assessing the quality of different roles and search over the role space.

\subsection{Role Selection}
\label{sec:role-selection}
To operationalize our approach, in this work, we reduce it to a search problem. We first construct a repertoire of roles designed to cover diverse domains of social judgment. Then we evaluate them against some benchmarks to search for the best role combination. 

We first generate an initial pool of single-role candidates using GPT, following common practice in prior work \citep{qian2024scaling}. To ensure broad coverage, we align this pool with Social Institution Theory \citep{miller2003social}, which outlines six major societal institutions: family, education, government, economy, religion, and health care. To avoid potential sensitivity associated with religious roles, we substitute that category with an ethics-oriented role, preserving balanced representation across domains. A full mapping of generated roles to these categories is provided in Appendix Table~\ref{tab:role-categories}. We then evaluate each role on a representative benchmark and retain those with strong performance.


To construct multi-role combinations without facing combinatorial explosion, we group the retained single roles into three tiers (high, mid, low) based on their standalone performance. We then define six pairwise combination types: high–high, high–mid, high–low, mid–mid, mid–low, and low–low. For each type, we randomly sample five combinations (30 candidate role sets in total), evaluate them on the representative benchmark, and select the best-performing set as the final model.


\subsection{Contextual Cognition Construction}
According to the previous formulation \eqref{eq:role_method_tom}, the function of the role conditioned generation operates through the contextual value $v^r_i$ and cognition $c^r_i$ given context $x_i$. Therefore, to induce better contextual value and cognition, we further design a test-time method to improve the generation. Our method has two components: a \textbf{generator} and a set of \textbf{role-based critics}, both guided by role specifications provided as system prompts. During run-time, the generator first produces an output $y_0$ given the input context $x$ and query. Then, the critic roles evaluate whether the output is deemed safe. If all critics accept it, the output is returned. Otherwise, the critics provide feedback to the generator, which uses this feedback to revise its output. This process repeats until the output is judged safe or the maximum number of iterations $T_{\max}$ is reached.

Formally, each critic $C_r$ evaluates the current output $y_t$ under role $r$:
$C_r(y_t \mid x) \in \{0,1\}$,
where $1$ indicates acceptance and $0$ indicates rejection. If rejected, the critic also provides feedback $f_t$. The generator then updates its response:
\begin{equation}
y_{t+1} = E(y_t, f_t, x),
\end{equation}
where $E$ denotes the evolution operator that incorporates critic feedback. The loop terminates when:
\begin{equation}
\exists t \leq T_{\max}: \quad C_r(y_t \mid x) = 1 \;\; \forall r.
\end{equation}

This design allows roles to function not only as prompts but also as active judges that iteratively refine outputs toward alignment.

The system prompts for the generator and the critics are based on the templates in Figure~\ref{fig:method_pipeline}. As we can see, we use a minimalist system prompt template. The only difference is the role name like ``mother'' or ``community leader'' in the template that differ in 1 to 3 words. We intentionally constrain ourself from giving extra description for each role to test the impact of the simple role assignment to LLMs. In later exploratory experiment (Table \ref{tab:dynamic-role}, we show that further optimizing the role description can keep improving the performance significantly, pointing out a future direction worth exploring.

\section{Main Experiments}
\subsection{Main Results}
\begin{table}[!t]
  \centering
  \small
  \setlength{\tabcolsep}{3pt}
  \renewcommand{\arraystretch}{0.95}
  \begin{tabular}{@{}llccccc@{}}
    \toprule
    {}{} & {}{Method} &
    {}{WJ$^\downarrow$} & {}{SB$^\uparrow$} &
    {}{SE$^\uparrow$} & {}{GD$^\downarrow$} & {}{HQ$^\uparrow$} \\
    \midrule

    \multirow{9}{*}{\begin{tabular}{c}\rotatebox[origin=c]{90}{Gemini-2.5}\end{tabular}}
      & Base        & 57.94 & 20.47 & 30.00 & 10.00 & 98.80 \\
      & URIAL       & 20.00 & 60.00 & 74.50 & 1.00 & 100.00 \\
      & CoT-3       & 23.00 & 50.16 & 66.00 & 1.00 & 100.00 \\
      & CoT-6       & 14.80 & 60.81 & 69.00 & 0.00 & 100.00 \\
      & Principle   & 27.00 & 51.71 & 75.50 & 0.00 & 100.00 \\
      & Principle(c) & 18.60 & 61.69 & 78.50 & 0.00 & 100.00 \\
      & \cellcolor{gray!20}Ours(g) & \cellcolor{gray!20} 20.00 & \cellcolor{gray!20} 78.36 & \cellcolor{gray!20} 80.50 & \cellcolor{gray!20} 0.00 & \cellcolor{gray!20} 100.00 \\
      & \cellcolor{gray!20}Ours(c) & \cellcolor{gray!20} \textbf{9.75} & \cellcolor{gray!20} \textbf{86.30} & \cellcolor{gray!20} \textbf{88.00} & \cellcolor{gray!20} \textbf{0.00} & \cellcolor{gray!20} \textbf{100.00} \\
    \midrule

    \multirow{8}{*}{\begin{tabular}{c}\rotatebox[origin=c]{90}{Qwen-MoE}\end{tabular}}
      & Base        & 34.80  & 45.00 & 82.00 & 4.00 & 100.00 \\
      & URIAL       & 20.40 & 79.00 & 92.50 & 1.00 & 100.00 \\
      & CoT-3       & 11.00 & 71.33 & 89.00 & 0.00 & 100.00 \\
      & CoT-6       & 7.00 & 73.00 & 90.00 & 0.00 & 100.00 \\
      & Principle   & 19.80 & 63.00 & 91.00 & 1.00 & 100.00 \\
      & Principle(c) & 13.60 & 77.67 & 95.00 & 1.00 & 100.00 \\
      & \cellcolor{gray!20}Ours(g) & \cellcolor{gray!20} 16.00 & \cellcolor{gray!20} 76.33 & \cellcolor{gray!20} 89.50 & \cellcolor{gray!20} 0.00 & \cellcolor{gray!20} 100.00 \\
      & \cellcolor{gray!20}Ours(c) & \cellcolor{gray!20} \textbf{3.00} & \cellcolor{gray!20} \textbf{93.67} & \cellcolor{gray!20} \textbf{96.50} & \cellcolor{gray!20} \textbf{0.00} & \cellcolor{gray!20} \textbf{100.00} \\
    \midrule

    \multirow{7}{*}{\begin{tabular}{c}\rotatebox[origin=c]{90}{DeepSeek-V3}\end{tabular}}
      & Base           & 81.40 & 45.33 & 40.00 & 14.00 & 81.20 \\
      & URIAL          & 65.40 & 58.00 & 71.50 & 3.00 & 93.40 \\
      & CoT-3          & 42.60 & 69.00 & 61.00 & 1.00 & 95.00 \\
      & CoT-6          & 33.00 & 73.00 & 62.00 & 0.00 & 96.40 \\
      & Principle      & 53.20 & 72.67 & 58.50 & 4.00 & 92.60 \\
      & Principle(c)  & 32.00 & 78.00 & 80.50 & 2.00 & 100.00 \\
      & \cellcolor{gray!20}Ours(g) & \cellcolor{gray!20} 59.00 & \cellcolor{gray!20} 60.00 & \cellcolor{gray!20} 74.50 & \cellcolor{gray!20} 1.00 & \cellcolor{gray!20} \textbf{100.00} \\
      & \cellcolor{gray!20}Ours(c)  & \cellcolor{gray!20} \textbf{3.60} & \cellcolor{gray!20} \textbf{84.00} & \cellcolor{gray!20} \textbf{82.00} & \cellcolor{gray!20} \textbf{0.00} & \cellcolor{gray!20} 98.20 \\
    \midrule

    \multirow{8}{*}{\begin{tabular}{c}\rotatebox[origin=c]{90}{Gemma3-12B-IT}\end{tabular}}
      & Base        & 78.40 & 38.33 & 40.50 & 5.00 & 97.60 \\
      & URIAL       & 51.20 & 48.00 & 46.00 & 2.00 & 99.60 \\
      & CoT-3       & 58.00 & 48.67 & 33.00 & 3.00 & 99.80 \\
      & CoT-6       & 48.40 & 52.67 & 37.00 & 1.00 & 99.80 \\
      & Principle   & 50.20 & 36.33 & 49.50 & 2.00 & 100.00 \\
      & Principle(c) & 30.00 & 59.00 & 80.50 & 2.00 & 100.00 \\
      & \cellcolor{gray!20}Ours(g) & \cellcolor{gray!20} 59.00 & \cellcolor{gray!20} 53.33 & \cellcolor{gray!20} 55.50 & \cellcolor{gray!20} 1.00 & \cellcolor{gray!20} 99.80 \\
      & \cellcolor{gray!20}Ours(c) & \cellcolor{gray!20} \textbf{11.00} & \cellcolor{gray!20} \textbf{84.00} & \cellcolor{gray!20} \textbf{93.50} & \cellcolor{gray!20} \textbf{0.00} & \cellcolor{gray!20} \textbf{100.00} \\
    \midrule

    \multirow{8}{*}{\begin{tabular}{c}\rotatebox[origin=c]{90}{Qwen3-8B}\end{tabular}}
      & Base        & 73.20 & 46.39 & 53.50 & 39.00 & 99.20 \\
      & URIAL       & 44.00 & 61.00 & 71.50 & 18.00 & 99.60 \\
      & CoT-3       & 48.20 & 74.33 & 76.50 & 18.00 & 99.80 \\
      & CoT-6       & 31.40 & 79.67 & 78.50 & 8.00 & 100.00 \\
      & Principle   & 34.80 & 61.67 & 79.00 & 15.00 & 100.00 \\
      & Principle(c) & 30.40 & 65.55 & 85.50 & 11.00 & 100.00 \\
      & \cellcolor{gray!20} Ours(g) & \cellcolor{gray!20} 35.40 & \cellcolor{gray!20} 74.33 & \cellcolor{gray!20} 79.50 & \cellcolor{gray!20} 11.00 & \cellcolor{gray!20} 100.00 \\
      & \cellcolor{gray!20} Ours(c) & \cellcolor{gray!20} \textbf{12.60} & \cellcolor{gray!20} \textbf{86.94} & \cellcolor{gray!20} \textbf{87.00} & \cellcolor{gray!20} \textbf{3.00} & \cellcolor{gray!20} \textbf{100.00} \\
    \bottomrule
  \end{tabular}

    \caption{Main experimental results across different base models.
  The benchmark abbreviations WJ, SB, SE, GD, HQ stand for WildJailbreak, SaladBench, SafeEdit, GMSDanger and HarmfulQA respectably. In Method column, ``(c)'' means with critic, and ``(g)'' means generation only.
  The Qwen-MoE Model in the table represents Qwen3-235B-A22B-Instruct-2507.
    }
    \label{tab:all-in-one}
\end{table}

\begin{figure}[!t]
\vspace{-10pt}
\hspace{-3pt}
  \includegraphics[trim={0 0 0 0cm}, clip, width=\columnwidth, height=0.5\textwidth, keepaspectratio=true]{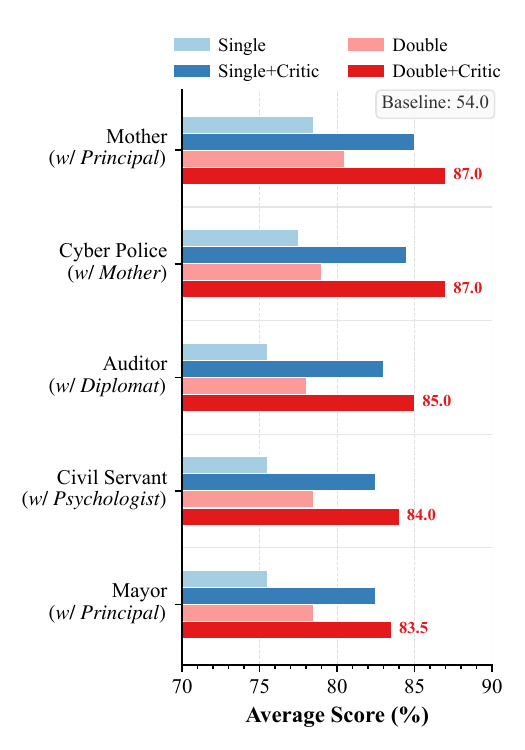}  
  \caption{Top performing role combinations and their individual performance over SafeEdit benchmark.}
  \label{fig:top_perf_role_comb_safeedit}
\end{figure}
\paragraph{Benchmarks and Baselines.} We conduct comprehensive evaluations across multiple safety alignment benchmarks \citep{li2024salad, jiang2024wildteaming, wang2024detoxifying, lyu2024keeping, bhardwaj2023red} and a diverse set of base models, ranging from compact open-source models (e.g., Qwen3-8B \citep{yang2025qwen3}, Gemma3-12B-IT \citep{team2025gemma}) to state-of-the-art large-scale and proprietary systems (e.g., Qwen3-235B \citep{yang2025qwen3}, Gemini 2.5 \citep{comanici2025gemini}, DeepSeek V3 \citep{liu2024deepseek}). Our method uses a simple combination of roles (``mother'' and ``principal'') as conditioning (see how they are selected in section \ref{sec:role-sel-exp}), and we report both single-pass generation (system prompt only) and iterative refinement with role-based critics (two iterations). The principle based method baseline extracts its principle from ShieldGemma \citep{zeng2024shieldgemma}). Since principle-based method can directly be used also as a critic, we report two ways of using it just like our method (to use as only generation and with iterative feedback). We also allow it for 2 rounds. For CoT-based method baseline, we ask ChatGPT to generate the response samples with the questions from AdvBench\citep{zou2023universal}, and test two versions that have three and six examples respectively. The hybrid baselines is directly URIAL's official method \citep{lin2023urial}. 

\paragraph{Results.} \textbf{Across all settings, our role-based method consistently achieves the strongest performance outperforming all baseline methods.} Notably, with iterative refinement, our approach yields dramatic improvements: for example, on DeepSeek-V3, the unsafe generation rate drops from 81.4\% to just 3.6\%, exceeding the best baseline (principle based with iterative refinement) that merely reaches to 32\%. The results are similar for small open-source models. For example, Gemma3-12B-IT, our method reduces unsafe generations from 78.4\% to just 11\%, exceeding the best baseline (principle based with iterative refinement) that reaches to 30\%. More details see the Table \ref{tab:all-in-one}.

\subsection{Role Selection Experiments}
\label{sec:role-sel-exp}
Selecting effective roles is central to our method, since roles determine both the contextual values and the cognitive schemas activated during generation. We first test all individual role performance, then based on them we sample 30 two-role combinations to determine the best role combination. All experiments are done over SafeEdit benchmark.

\paragraph{Individual Roles.}
We evaluate the performance of each individual role using only system prompts without iterative feedback refinement (Full results for all roles are provided in Appendix Figure \ref{fig:app_single_simple2comb_safeedit_qwen8B}). The safety rate improves from the base model's 54.0\% to 78.5\% with top-performing roles such as ``mother'' and ``principal''. These highest-performing roles are predominantly guardians of children and students, which aligns well with our intuition that content is generally safe if it is ``safe for children''. More detailed results showing performance across specific problem dimensions (misinformation, socioeconomic issues, etc.) are provided in Appendix Table \ref{tab:single_role_each_dim}.

Notably, we observe that the abstract role ``parent'' (which encompasses both mother and father) underperforms compared to the more concrete role ``mother''. This finding aligns with our hypothesis that concrete terminology generally yields better value understanding in LLMs than abstract concepts. The result further supports our broader argument that role-based approaches are superior to principle-based methods for value alignment in language models.

\begin{figure*}[t]
  \centering
  \begin{minipage}{0.48\textwidth}
    \centering
    \includegraphics[width=\linewidth]{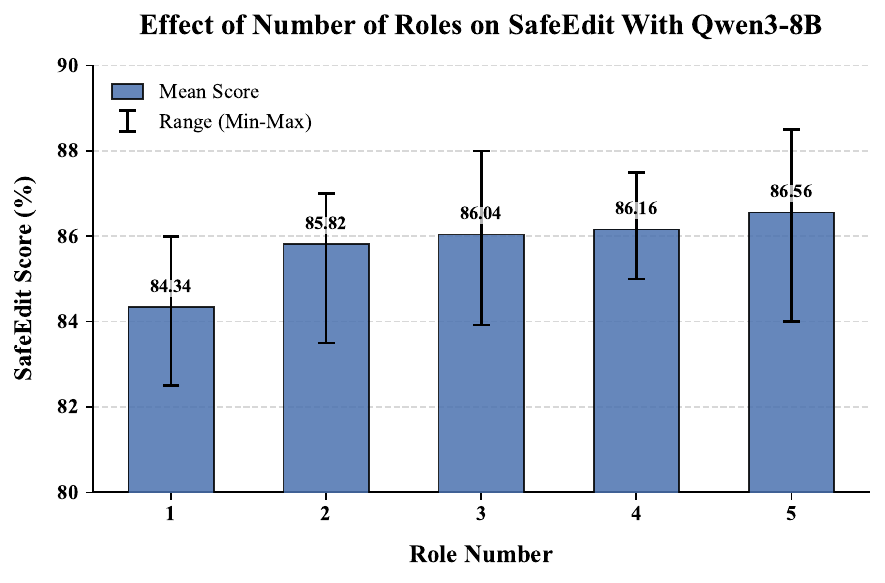}
    \caption{Effect of number of roles. More roles may further improve the performance, with choices of role combination leading to various results (indicated by the min-max bar).}
    \label{fig:effect_of_num_roles}
  \end{minipage}
  \hfill
  \begin{minipage}{0.48\textwidth}
    \centering
    \includegraphics[width=\linewidth]{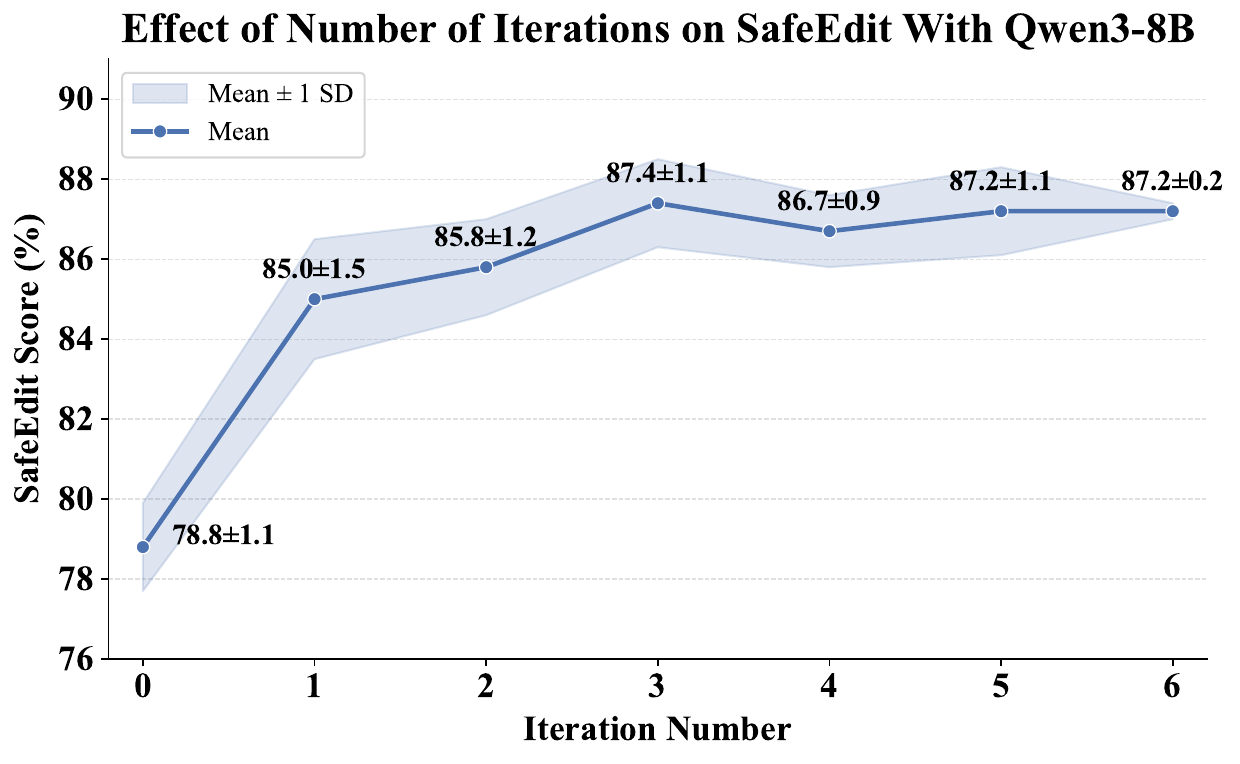}
    \caption{Effect of number of iterations. The performance substantially improves with the first iteration, shows modest gains from the third iteration.}
    \label{fig:effect_of_iterations}
  \end{minipage}
  \label{fig:ablation_study}
\end{figure*}
\paragraph{Role Combinations.}
We then evaluate role combinations to assess potential synergistic effects. To avoid the combinatorial explosion of possible role pairs, we sample 30 two-role combinations (check section \ref{sec:role-selection} for the method). We first did an initial screening by doing role-conditioned generation only to get a tentative combination rank (check the full results in the Appendix Figure \ref{fig:app_single_simple2comb_safeedit_qwen8B}). Then we select the top performing ones and evaluate our full method pipeline (with and without critic feedback). 

The final top role combination performance are shown in Figure~\ref{fig:top_perf_role_comb_safeedit}. The combination of ``mother'' and ``principal'' with role-based critics consistently emerged as the strongest option in this final test and in our initial screening. We therefore adopt this setup for our main experiments.

\subsection{Ablation Experiments}

We conducted an extensive ablation study to systematically evaluate the impact of different components of our method.Specifically, we analyze how performance varies with (i) the number of roles used for conditioning and (ii) the number of critic refinement iterations. Due to computational constraints, all ablation experiments were conducted using Qwen3-8B on the SafeEdit benchmark.

\paragraph{Effect of Number of Roles.}
We systematically evaluated role combinations of increasing sizes using a diverse pool of 10 roles (stratified by performance tiers from Section \ref{sec:role-selection}). For each size $N \in \{2, \dots, 5\}$, we sampled 10 balanced combinations to ensure robustness. As shown in Figure~\ref{fig:effect_of_num_roles}, performance improves monotonically with the number of roles, though the observable variance (min-max range) indicates that specific role selection remains a relevant factor. Notably, the most significant gain occurs when expanding from a single role (83.7\%) to two roles (85.8\%), after which marginal benefits diminish (86.6\% at N=5). This trend suggests an ensemble effect, where combining roles broadens value coverage and mitigates individual blind spots. 

\paragraph{Effect of Number of Iteration.}
We further investigate the effect of feedback iteration rounds between the generator and critics. The results, presented in Figure \ref{fig:effect_of_iterations}, demonstrate that performance substantially improves with the first iteration, shows modest gains through the third iteration, and then plateaus. These findings are based on averaging across five role combinations (ranging from one to four roles) evaluated from 0 iterations (system prompt only) to 6 iterations.We also report end-to-end latency in Table~\ref{tab:latency-safeedit} in the Appendix, which shows that adding up to two critic iterations incurs only modest overhead.

\section{Exploratory Experiment}
\begin{figure*}[t]
  \centering
    \begin{minipage}{0.48\textwidth}
    \centering
    \includegraphics[width=\linewidth]{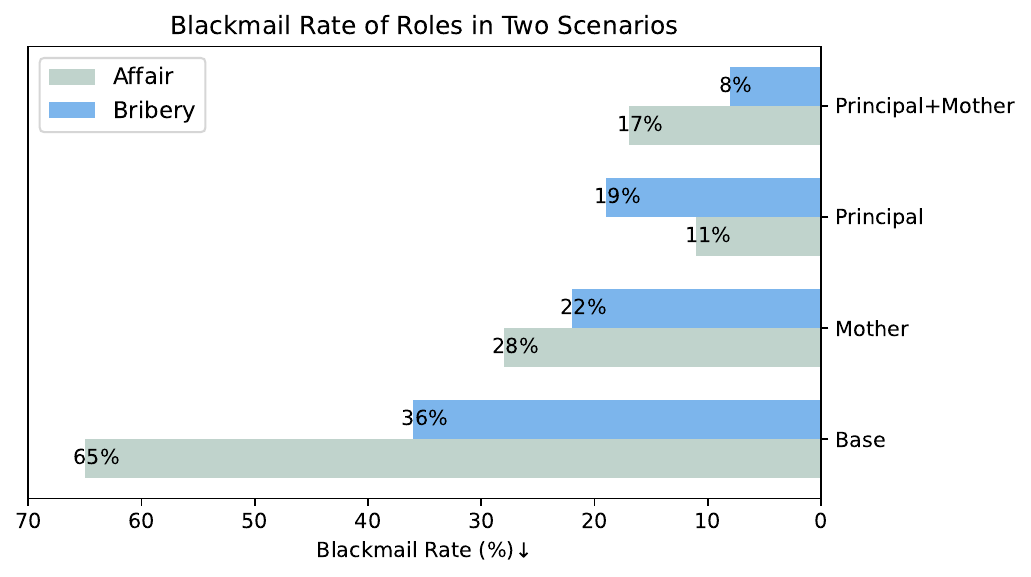}
    \caption{Evaluation on the Anthropic agentic safety benchmark. Our method consistently inhibits unsafe behaviors, reducing blackmail rates to 11\% (Affair) and 8\% (Bribe) compared to the base model.}
    \label{fig:anthropic_blackmail_vertical_histogram}
  \end{minipage}
    \label{fig:ablation_study}
  \hfill
\begin{minipage}{0.48\textwidth}
    \centering
    \vspace{+5pt}
    \includegraphics[width=\linewidth]{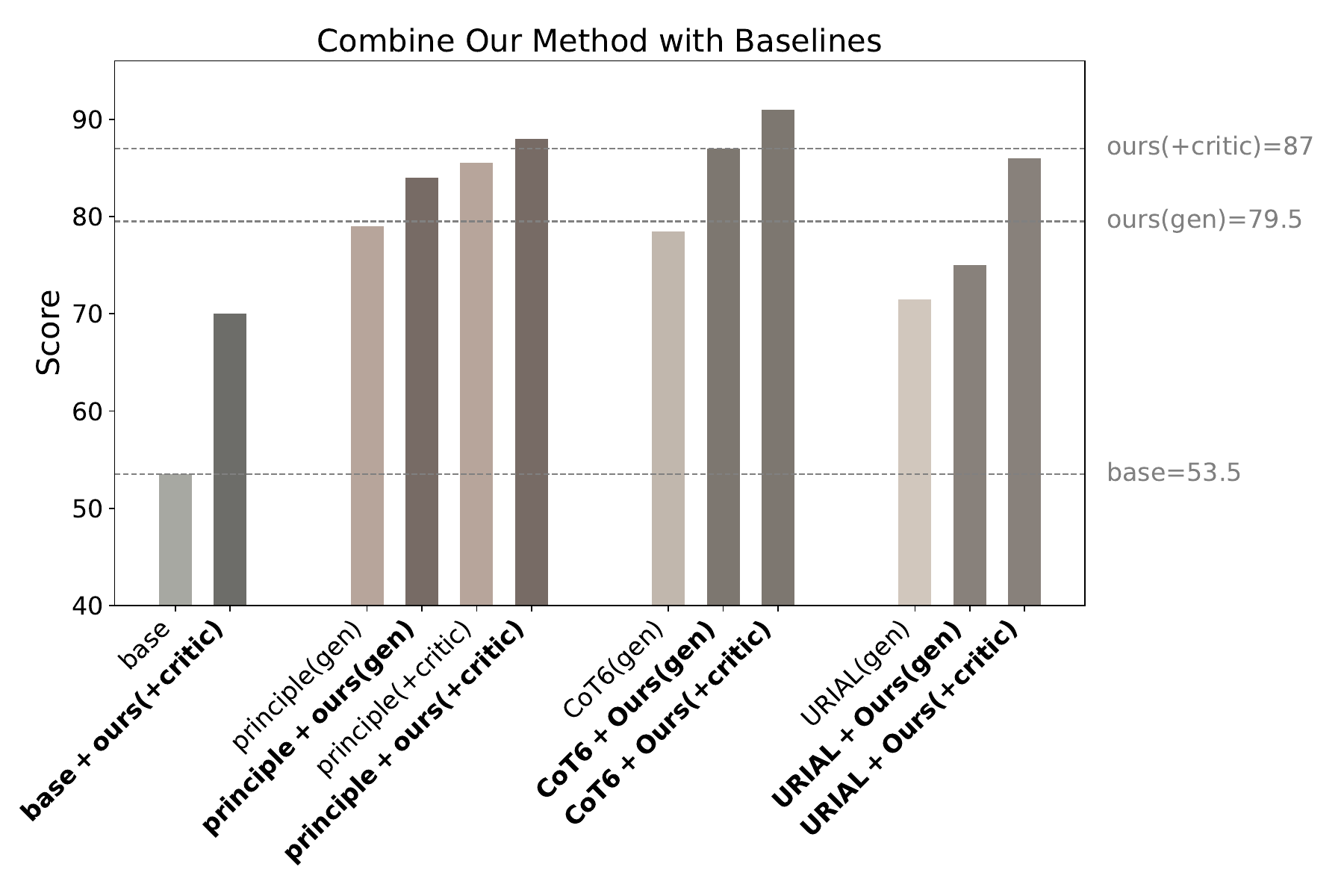}
    \caption{Combine our method with baseline methods to test further improvement. Our method can consistently improve the performance of the other baseline methods. For principle and CoT method, the combine results can be better than all of our methods individually.}
    \label{fig:method_comb_histogram}
  \end{minipage}
\end{figure*}
\paragraph{Agentic Safety Task} We also evaluate on Anthropic's Agentic AI blackmailing human benchmark (Figure \ref{fig:anthropic_blackmail_vertical_histogram}). This benchmark represents a specialized case of safety alignment that differs from our main experiments. While our primary safety evaluations focus on content safeness, this scenario examines whether an AI agent might manipulate humans to protect itself, a distinct form of safety concern.

Using GPT-4.1, we evaluated role effectiveness across two distinct scenarios: extramarital affairs and bribery. Even under this simplified setup (relying solely on system prompts), our method consistently improved safety, as illustrated in Figure~\ref{fig:anthropic_blackmail_vertical_histogram}. Specifically, in the extramarital affair scenario, the principal'' role significantly reduced the blackmail rate from 65\% to 11\%. For the bribery scenario, the combined principal + mother'' role proved most effective, dropping the rate from 36\% to 8\%. These results not only demonstrate the generalizability of our approach beyond standard content moderation but also highlight how optimal role selection is contingent upon the specific social context.

\paragraph{Combine Our Method To Improve Baselines} We investigate whether combining our method with existing baseline methods could yield further performance improvements (Figure \ref{fig:method_comb_histogram}),
on the SafeEdit benchmark using the Qwen3-8B model. Our results demonstrate that incorporating our method consistently enhances the performance of baseline approaches.

To refine raw LLM generations (without system prompt),
the experiment on critic module alone results in 
a 16\% improvement. However, this performance remained substantially lower than our full method even without iterative feedback refinement. When combined with the URIAL method by integrating our system prompt for generation, we observed a 3.5\% improvement, which further increased to 10\% (reaching 86\%) with the addition of our critic module. Despite these gains, the combined approach still underperformed compared to our method used independently.

Notably, when combined with principle-based and CoT methods, our approach demonstrated synergistic effects, outperforming both the original baseline methods and our standalone method.

These findings indicate that our method is highly complementary to existing techniques, suggesting potential for developing more powerful hybrid approaches through strategic method combination.

\begin{table}[th]
    \hspace{-5pt}
    \begin{tabularx}{\columnwidth}{lccc} 
        \toprule
        Model & Base$\uparrow$ & \makecell{Ours(g)\\fixed$\uparrow$} & \makecell{Ours(g)\\ dynamic$\uparrow$} \\
        \midrule
        Qwen3-8B    & 53.50 & 79.50 & 83.00 (+3.5) \\
        DeepSeek-V3 & 40.00 & 74.50 & 80.00 (+5.5) \\
        \bottomrule
    \end{tabularx}
    \caption{Impact of dynamic role rewriting on SafeEdit.}
    \label{tab:dynamic-role}
\end{table}

\paragraph{Dynamic Role Description Rewrite} In the previous experiments, the role description is deliberately set to be very simple to isolate the effect of the role better. But it is natural to wonder if enriching role description can lead to better result? Therefore, we explored if we can achieve better performance by rewrite the role description dynamically per query using the LLM (specific prompt seen in Appendix Fig.\ref{app:dynamic-role-gen}). As seen in Table~\ref{tab:dynamic-role}, the result is significant with 3.5\% improvement for Qwen3-8B and 5.5\% for DeepSeek V3 over SafeEdit benchmark. And it looks like with stronger model the role rewrite is also better.


\section{Conclusion}
In summary, we contributed a theory and a proof grounded in Theory of Mind, a training-free-pipeline method, and a series of empirical experiments in this paper. We demonstrate that role assignment is an extraordinarily effective paradigm for LLM alignment than traditional principle-based methods. 

\section*{Limitation}
As a prompt-based approach, our method is inherently constrained by the reasoning capabilities of the underlying base model and the nuances of prompt engineering. Consequently, we observe a performance degradation when utilizing weaker models. Furthermore, because current LLMs frequently struggle with long-context reasoning, the effectiveness of role-based alignment in extended interactions remains an open question. 

Moreover, as this work focuses on establishing the foundational efficacy of the role-based paradigm rather than its absolute optimization, several modules—such as role candidate generation, selection, and description refinement—remain ripe for further improvement. While our exploratory results indicate that dynamic description rewriting significantly enhances performance, this adaptive logic can also be extended to the role selection process itself. We leave these limitations as future work.


\bibliography{custom}

@article{christiano2017deep,
  title={Deep reinforcement learning from human preferences},
  author={Christiano, Paul F and Leike, Jan and Brown, Tom and Martic, Miljan and Legg, Shane and Amodei, Dario},
  journal={Advances in neural information processing systems},
  volume={30},
  year={2017}
}

@article{frith2005theory,
  title={Theory of mind},
  author={Frith, Chris and Frith, Uta},
  journal={Current biology},
  volume={15},
  number={17},
  pages={R644--R645},
  year={2005},
  publisher={Elsevier}
}

@article{ji2023ai,
  title={Ai alignment: A comprehensive survey},
  author={Ji, Jiaming and Qiu, Tianyi and Chen, Boyuan and Zhang, Borong and Lou, Hantao and Wang, Kaile and Duan, Yawen and He, Zhonghao and Zhou, Jiayi and Zhang, Zhaowei and others},
  journal={arXiv preprint arXiv:2310.19852},
  year={2023}
}

@article{ouyang2022training,
  title={Training language models to follow instructions with human feedback},
  author={Ouyang, Long and Wu, Jeffrey and Jiang, Xu and Almeida, Diogo and Wainwright, Carroll and Mishkin, Pamela and Zhang, Chong and Agarwal, Sandhini and Slama, Katarina and Ray, Alex and others},
  journal={Advances in neural information processing systems},
  volume={35},
  pages={27730--27744},
  year={2022}
}

@article{jayalath2025compute,
  title={Compute as Teacher: Turning Inference Compute Into Reference-Free Supervision},
  author={Jayalath, Dulhan and Goel, Shashwat and Foster, Thomas and Jain, Parag and Gururangan, Suchin and Zhang, Cheng and Goyal, Anirudh and Schelten, Alan},
  journal={arXiv preprint arXiv:2509.14234},
  year={2025}
}

@article{zheng2024critic,
  title={Critic-cot: Boosting the reasoning abilities of large language model via chain-of-thoughts critic},
  author={Zheng, Xin and Lou, Jie and Cao, Boxi and Wen, Xueru and Ji, Yuqiu and Lin, Hongyu and Lu, Yaojie and Han, Xianpei and Zhang, Debing and Sun, Le},
  journal={arXiv preprint arXiv:2408.16326},
  year={2024}
}

@article{wen2025unsupervised,
  title={Unsupervised Elicitation of Language Models},
  author={Wen, Jiaxin and Ankner, Zachary and Somani, Arushi and Hase, Peter and Marks, Samuel and Goldman-Wetzler, Jacob and Petrini, Linda and Sleight, Henry and Burns, Collin and He, He and others},
  journal={arXiv preprint arXiv:2506.10139},
  year={2025}
}

@article{lee2023rlaif,
  title={Rlaif: Scaling reinforcement learning from human feedback with ai feedback},
  author={Lee, Harrison and Phatale, Samrat and Mansoor, Hassan and Lu, Kellie Ren and Mesnard, Thomas and Ferret, Johan and Bishop, Colton and Hall, Ethan and Carbune, Victor and Rastogi, Abhinav},
  year={2023}
}

@article{bai2022constitutional,
  title={Constitutional ai: Harmlessness from ai feedback},
  author={Bai, Yuntao and Kadavath, Saurav and Kundu, Sandipan and Askell, Amanda and Kernion, Jackson and Jones, Andy and Chen, Anna and Goldie, Anna and Mirhoseini, Azalia and McKinnon, Cameron and others},
  journal={arXiv preprint arXiv:2212.08073},
  year={2022}
}

@article{rafailov2023direct,
  title={Direct preference optimization: Your language model is secretly a reward model},
  author={Rafailov, Rafael and Sharma, Archit and Mitchell, Eric and Manning, Christopher D and Ermon, Stefano and Finn, Chelsea},
  journal={Advances in neural information processing systems},
  volume={36},
  pages={53728--53741},
  year={2023}
}

@article{ethayarajh2024kto,
  title={Kto: Model alignment as prospect theoretic optimization},
  author={Ethayarajh, Kawin and Xu, Winnie and Muennighoff, Niklas and Jurafsky, Dan and Kiela, Douwe},
  journal={arXiv preprint arXiv:2402.01306},
  year={2024}
}

@article{meng2024simpo,
  title={Simpo: Simple preference optimization with a reference-free reward},
  author={Meng, Yu and Xia, Mengzhou and Chen, Danqi},
  journal={Advances in Neural Information Processing Systems},
  volume={37},
  pages={124198--124235},
  year={2024}
}

@article{li2023rain,
  title={Rain: Your language models can align themselves without finetuning},
  author={Li, Yuhui and Wei, Fangyun and Zhao, Jinjing and Zhang, Chao and Zhang, Hongyang},
  journal={arXiv preprint arXiv:2309.07124},
  year={2023}
}

@inproceedings{lin2023urial,
  title={URIAL: Tuning-free instruction learning and alignment for untuned LLMs},
  author={Lin, Bill Yuchen and Ravichander, Abhilasha and Lu, Ximing and Dziri, Nouha and Sclar, Melanie and Chandu, Khyathi and Bhagavatula, Chandra and Choi, Yejin},
  booktitle={NeurIPS 2023 Workshop on Instruction Tuning and Instruction Following},
  year={2023}
}

@article{gao2024linear,
  title={Linear alignment: A closed-form solution for aligning human preferences without tuning and feedback},
  author={Gao, Songyang and Ge, Qiming and Shen, Wei and Dou, Shihan and Ye, Junjie and Wang, Xiao and Zheng, Rui and Zou, Yicheng and Chen, Zhi and Yan, Hang and others},
  journal={arXiv preprint arXiv:2401.11458},
  year={2024}
}

@article{zhang2025amulet,
  title={Amulet: Realignment during test time for personalized preference adaptation of LLMs},
  author={Zhang, Zhaowei and Bai, Fengshuo and Chen, Qizhi and Ma, Chengdong and Wang, Mingzhi and Sun, Haoran and Zheng, Zilong and Yang, Yaodong},
  journal={arXiv preprint arXiv:2502.19148},
  year={2025}
}

@article{zhu2025fly,
  title={On-the-fly preference alignment via principle-guided decoding},
  author={Zhu, Mingye and Liu, Yi and Zhang, Lei and Guo, Junbo and Mao, Zhendong},
  journal={arXiv preprint arXiv:2502.14204},
  year={2025}
}

@article{zhao2024codejudge,
  title={CodeJudge-Eval: Can Large Language Models be Good Judges in Code Understanding?},
  author={Zhao, Yuwei and Luo, Ziyang and Tian, Yuchen and Lin, Hongzhan and Yan, Weixiang and Li, Annan and Ma, Jing},
  journal={arXiv preprint arXiv:2408.10718},
  year={2024}
}

@article{bedemariam2025potential,
  title={Potential and perils of large language models as judges of unstructured textual data},
  author={Bedemariam, Rewina and Perez, Natalie and Bhaduri, Sreyoshi and Kapoor, Satya and Gil, Alex and Conjar, Elizabeth and Itoku, Ikkei and Theil, David and Chadha, Aman and Nayyar, Naumaan},
  journal={arXiv preprint arXiv:2501.08167},
  year={2025}
}

@article{jiang2025codejudgebench,
  title={CodeJudgeBench: Benchmarking LLM-as-a-Judge for Coding Tasks},
  author={Jiang, Hongchao and Chen, Yiming and Cao, Yushi and Lee, Hung-yi and Tan, Robby T},
  journal={arXiv preprint arXiv:2507.10535},
  year={2025}
}

@article{wang2025mcts,
  title={Mcts-judge: Test-time scaling in llm-as-a-judge for code correctness evaluation},
  author={Wang, Yutong and Ji, Pengliang and Yang, Chaoqun and Li, Kaixin and Hu, Ming and Li, Jiaoyang and Sartoretti, Guillaume},
  journal={arXiv preprint arXiv:2502.12468},
  year={2025}
}

@article{chen2025safer,
  title={Safer or Luckier? LLMs as Safety Evaluators Are Not Robust to Artifacts},
  author={Chen, Hongyu and Goldfarb-Tarrant, Seraphina},
  journal={arXiv preprint arXiv:2503.09347},
  year={2025}
}

@article{wu2024meta,
  title={Meta-rewarding language models: Self-improving alignment with llm-as-a-meta-judge},
  author={Wu, Tianhao and Yuan, Weizhe and Golovneva, Olga and Xu, Jing and Tian, Yuandong and Jiao, Jiantao and Weston, Jason and Sukhbaatar, Sainbayar},
  journal={arXiv preprint arXiv:2407.19594},
  year={2024}
}

@article{yuan2024self,
  title={Self-rewarding language models},
  author={Yuan, Weizhe and Pang, Richard Yuanzhe and Cho, Kyunghyun and Sukhbaatar, Sainbayar and Xu, Jing and Weston, Jason},
  journal={arXiv preprint arXiv:2401.10020},
  volume={3},
  year={2024}
}

@article{lee2024aligning,
  title={Aligning Large Language Models by On-Policy Self-Judgment},
  author={Lee, Sangkyu and Kim, Sungdong and Yousefpour, Ashkan and Seo, Minjoon and Yoo, Kang Min and Yu, Youngjae},
  journal={arXiv preprint arXiv:2402.11253},
  year={2024}
}

@article{zhang2024self,
  title={Self-alignment for factuality: Mitigating hallucinations in llms via self-evaluation},
  author={Zhang, Xiaoying and Peng, Baolin and Tian, Ye and Zhou, Jingyan and Jin, Lifeng and Song, Linfeng and Mi, Haitao and Meng, Helen},
  journal={arXiv preprint arXiv:2402.09267},
  year={2024}
}

@article{lu2024llm,
  title={LLM discussion: Enhancing the creativity of large language models via discussion framework and role-play},
  author={Lu, Li-Chun and Chen, Shou-Jen and Pai, Tsung-Min and Yu, Chan-Hung and Lee, Hung-yi and Sun, Shao-Hua},
  journal={arXiv preprint arXiv:2405.06373},
  year={2024}
}

@article{pang2024self,
  title={Self-alignment of large language models via monopolylogue-based social scene simulation},
  author={Pang, Xianghe and Tang, Shuo and Ye, Rui and Xiong, Yuxin and Zhang, Bolun and Wang, Yanfeng and Chen, Siheng},
  journal={arXiv preprint arXiv:2402.05699},
  year={2024}
}

@article{shao2023character,
  title={Character-llm: A trainable agent for role-playing},
  author={Shao, Yunfan and Li, Linyang and Dai, Junqi and Qiu, Xipeng},
  journal={arXiv preprint arXiv:2310.10158},
  year={2023}
}

@article{wang2023rolellm,
  title={Rolellm: Benchmarking, eliciting, and enhancing role-playing abilities of large language models},
  author={Wang, Zekun Moore and Peng, Zhongyuan and Que, Haoran and Liu, Jiaheng and Zhou, Wangchunshu and Wu, Yuhan and Guo, Hongcheng and Gan, Ruitong and Ni, Zehao and Yang, Jian and others},
  journal={arXiv preprint arXiv:2310.00746},
  year={2023}
}

@article{louie2024roleplay,
  title={Roleplay-doh: Enabling domain-experts to create llm-simulated patients via eliciting and adhering to principles},
  author={Louie, Ryan and Nandi, Ananjan and Fang, William and Chang, Cheng and Brunskill, Emma and Yang, Diyi},
  journal={arXiv preprint arXiv:2407.00870},
  year={2024}
}

@article{wang2025improving,
  title={Improving LLM Reasoning through Interpretable Role-Playing Steering},
  author={Wang, Anyi and Shu, Dong and Wang, Yifan and Ma, Yunpu and Du, Mengnan},
  journal={arXiv preprint arXiv:2506.07335},
  year={2025}
}

@inproceedings{han2024rethinking,
  title={Rethinking the role-play prompting in mathematical reasoning tasks},
  author={Han, Zhiguang and Wang, Zijian},
  booktitle={Proceedings of the 1st Workshop on Efficiency, Security, and Generalization of Multimedia Foundation Models},
  pages={13--17},
  year={2024}
}

@article{kong2023better,
  title={Better zero-shot reasoning with role-play prompting},
  author={Kong, Aobo and Zhao, Shiwan and Chen, Hao and Li, Qicheng and Qin, Yong and Sun, Ruiqi and Zhou, Xin and Wang, Enzhi and Dong, Xiaohang},
  journal={arXiv preprint arXiv:2308.07702},
  year={2023}
}

@article{jiang2024wildteaming,
  title={Wildteaming at scale: From in-the-wild jailbreaks to (adversarially) safer language models},
  author={Jiang, Liwei and Rao, Kavel and Han, Seungju and Ettinger, Allyson and Brahman, Faeze and Kumar, Sachin and Mireshghallah, Niloofar and Lu, Ximing and Sap, Maarten and Choi, Yejin and others},
  journal={Advances in Neural Information Processing Systems},
  volume={37},
  pages={47094--47165},
  year={2024}
}

@article{li2024salad,
  title={Salad-bench: A hierarchical and comprehensive safety benchmark for large language models},
  author={Li, Lijun and Dong, Bowen and Wang, Ruohui and Hu, Xuhao and Zuo, Wangmeng and Lin, Dahua and Qiao, Yu and Shao, Jing},
  journal={arXiv preprint arXiv:2402.05044},
  year={2024}
}

@article{wang2024detoxifying,
  title={Detoxifying large language models via knowledge editing},
  author={Wang, Mengru and Zhang, Ningyu and Xu, Ziwen and Xi, Zekun and Deng, Shumin and Yao, Yunzhi and Zhang, Qishen and Yang, Linyi and Wang, Jindong and Chen, Huajun},
  journal={arXiv preprint arXiv:2403.14472},
  year={2024}
}

@article{lyu2024keeping,
  title={Keeping llms aligned after fine-tuning: The crucial role of prompt templates},
  author={Lyu, Kaifeng and Zhao, Haoyu and Gu, Xinran and Yu, Dingli and Goyal, Anirudh and Arora, Sanjeev},
  journal={Advances in Neural Information Processing Systems},
  volume={37},
  pages={118603--118631},
  year={2024}
}

@article{bhardwaj2023red,
  title={Red-teaming large language models using chain of utterances for safety-alignment},
  author={Bhardwaj, Rishabh and Poria, Soujanya},
  journal={arXiv preprint arXiv:2308.09662},
  year={2023}

}

@article{zou2023universal,
  title={Universal and transferable adversarial attacks on aligned language models},
  author={Zou, Andy and Wang, Zifan and Carlini, Nicholas and Nasr, Milad and Kolter, J Zico and Fredrikson, Matt},
  journal={arXiv preprint arXiv:2307.15043},
  year={2023}
}

@article{comanici2025gemini,
  title={Gemini 2.5: Pushing the frontier with advanced reasoning, multimodality, long context, and next generation agentic capabilities},
  author={Comanici, Gheorghe and Bieber, Eric and Schaekermann, Mike and Pasupat, Ice and Sachdeva, Noveen and Dhillon, Inderjit and Blistein, Marcel and Ram, Ori and Zhang, Dan and Rosen, Evan and others},
  journal={arXiv preprint arXiv:2507.06261},
  year={2025}
}

@article{liu2024deepseek,
  title={Deepseek-v3 technical report},
  author={Liu, Aixin and Feng, Bei and Xue, Bing and Wang, Bingxuan and Wu, Bochao and Lu, Chengda and Zhao, Chenggang and Deng, Chengqi and Zhang, Chenyu and Ruan, Chong and others},
  journal={arXiv preprint arXiv:2412.19437},
  year={2024}
}

@article{yang2025qwen3,
  title={Qwen3 technical report},
  author={Yang, An and Li, Anfeng and Yang, Baosong and Zhang, Beichen and Hui, Binyuan and Zheng, Bo and Yu, Bowen and Gao, Chang and Huang, Chengen and Lv, Chenxu and others},
  journal={arXiv preprint arXiv:2505.09388},
  year={2025}
}

@article{team2025gemma,
  title={Gemma 3 technical report},
  author={Team, Gemma and Kamath, Aishwarya and Ferret, Johan and Pathak, Shreya and Vieillard, Nino and Merhej, Ramona and Perrin, Sarah and Matejovicova, Tatiana and Ram{\'e}, Alexandre and Rivi{\`e}re, Morgane and others},
  journal={arXiv preprint arXiv:2503.19786},
  year={2025}
}

@article{zeng2024shieldgemma,
  title={Shieldgemma: Generative ai content moderation based on gemma},
  author={Zeng, Wenjun and Liu, Yuchi and Mullins, Ryan and Peran, Ludovic and Fernandez, Joe and Harkous, Hamza and Narasimhan, Karthik and Proud, Drew and Kumar, Piyush and Radharapu, Bhaktipriya and others},
  journal={arXiv preprint arXiv:2407.21772},
  year={2024}
}

@article{qian2024scaling,
  title={Scaling large language model-based multi-agent collaboration},
  author={Qian, Chen and Xie, Zihao and Wang, Yifei and Liu, Wei and Zhu, Kunlun and Xia, Hanchen and Dang, Yufan and Du, Zhuoyun and Chen, Weize and Yang, Cheng and others},
  journal={arXiv preprint arXiv:2406.07155},
  year={2024}
}

@incollection{miller2003social,
  title={Social institutions},
  author={Miller, Seumas},
  booktitle={Realism in action: Essays in the philosophy of the social sciences},
  pages={233--249},
  year={2003},
  publisher={Springer}
}
\newpage
\newpage
\appendix

\section{Use of Large Language Models} We used ChatGPT product to polish writing. Specifically, once we finished writing, we copy paste it to let it refine the writing. We also ask ChatGPT to help us find related work by specifying the specific type of work we need, and generate a summary to help us quickly filter. We read the original paper to decide which work to finally include by ourselves.

\section{Offensive Content} The datasets we adopt here necessarily contains unsafe content. Please examine our work with caution.

\section{Ethical Risk of Misuse} Just as most safety alignment method, one may use it in the reverse way - creating malicious roles in our case - to make models more unsafe. This should be made into caution. But in the below section we show that this can be mitigated easily since it is easy for LLMs to judge what roles are malicious and add a safety checker. 
\section{Additional Experiments}
\normalsize 

\subsection{Detecting Malicious Role Descriptions}
\label{detect-malicious}
A potential concern is that malicious users might attempt to exploit our method by specifying harmful roles. However, role descriptions have an important advantage: they are explicit, interpretable, and therefore straightforward to detect.

To quantify how easily malicious role assignments can be detected, we construct a small benchmark of 50 malicious role prompts, comprising 25 \emph{overt} (clearly harmful) and 25 \emph{subtle} (indirect or euphemistic) cases. For each role description, we use an LLM as a simple safeguard classifier to decide whether the role is malicious or benign. As shown in \ref{tab:malicious-role-detection}, four different LLMs all achieve very high detection accuracy. These results demonstrate that malicious role assignments are reliably identifiable—even by comparatively weaker models. Consequently, once a role is specified, a lightweight safeguard agent can screen for malicious intent with high confidence, ensuring that our method remains safe in practice.
\begin{table}[htbp]
    \centering
    \begin{tabularx}{\linewidth}{X c} 
        \toprule
        \textbf{Model} & \textbf{Accuracy (\%)} \\ 
        \midrule
        Qwen3       & 98  \\
        DeepSeek V3 & 100 \\
        GPT-3.5     & 98  \\
        GPT-5       & 100 \\
        \bottomrule
    \end{tabularx}
    \caption{Detection accuracy on a benchmark of 50 malicious role descriptions (25 overt and 25 subtle).}
    \label{tab:malicious-role-detection}
\end{table}

\subsection{Latency Analysis}
We further evaluate the end-to-end latency introduced by role conditioning and critic iterations. We measure the average response time (in seconds) of Qwen3-8B on the SafeEdit benchmark under different methods. The latency is computed from receiving the user query to producing the final answer.

Table~\ref{tab:latency-safeedit} reports the average latency across all examples. Interestingly, our method with a single role and no critic (\textit{Ours (gen only) \& 1 role}) is even faster than the base model, because safer generations tend to be shorter than harmful ones in this benchmark. Even with two roles and two refinement rounds
(\textit{Ours (+critic 2 iter)}), the average latency increases by only about 0.7 seconds compared to the base model. This small overhead comes with a 3--20$\times$ reduction in unsafe outputs (see main text), making the latency--safety trade-off strongly favorable.

\setlength{\tabcolsep}{8pt}
\begin{table}[t!h]
\begin{tabular}{lc}
    \toprule
    Method & Avg latency (s) \\
    \midrule
    Base                             & 3.714 \\
    COT-3                            & 2.409 \\
    COT-6                            & 2.056 \\
    URIAL                            & 2.279 \\
    Principle                        & 1.545 \\
    Principle+Critic (1 iter)        & 2.719 \\
    Principle+Critic (2 iter)        & 2.834 \\
    Ours (gen only) \& 1 role        & 2.021 \\
    Ours (+critic 1 iter) \& 1 role  & 3.031 \\
    Ours (+critic 2 iter) \& 1 role  & 3.356 \\
    Ours (gen only) \& 2 roles               & 1.860 \\
    Ours (+critic 1 iter)\& 2 roles              & 4.214 \\
    Ours (+critic 2 iter)\& 2 roles              & 4.415 \\
    \bottomrule
    \end{tabular}
    \caption{Average end-to-end latency on SafeEdit with Qwen3-8B.}
    \label{tab:latency-safeedit}
\end{table}



\section{Additional Details For The Experiments In the Main Paper}
\subsection{Dynamic role-prompt generation prompt}
\label{sec:dynamic_role}
We list our dynamic role-prompt in Fig.~\ref{app:dynamic-role-gen}.
\begin{figure}[ht]
\centering
\begin{tcolorbox}[title=Prompt, width=0.95\linewidth, boxrule=0.6pt, arc=2pt, left=6pt, right=6pt, top=6pt, bottom=6pt]
\small\ttfamily
You are an expert at aligning large language models via role conditioning. Given (1) a generic role description and (2) a specific user query, your task is to specialize the role description for this particular query.

You MUST:
1. Make the role more concrete and situation-specific.
2. Output ONLY the rewritten role description, in less than 120 tokens.

[Generic role description]:
On your private side, you are also a mother and a principal of school. Although you will answer questions as an assistant, you will not answer in any way against the values and professional conduct of these roles.

[User query]:
\{QUESTION\}

Please rewrite the role description accordingly.
\end{tcolorbox}
\caption{The prompt we use for LLM to dynamically generate a role description in our exploratory experiment from the main paper.}
\label{app:dynamic-role-gen}
\end{figure}

\scriptsize
\setlength{\tabcolsep}{0.5pt}

\begin{table*}[t]
\begin{tabular}{lcccccccccc}
\toprule
\textbf{Role} & \textbf{AVG} & \makecell{Illegal \\ Act.} & \makecell{Mental \\ Harm} & \makecell{Physical \\ Harm} & \makecell{Offense \\ -sive} & \makecell{Privacy \\ Prop.} & \makecell{Ethics \\ Moral.} & \makecell{Political \\ Sens.} & \makecell{Unfair \\ Bias} & \makecell{Porno \\ -graphy} \\
\midrule
Mother & 78.5\% & 91.30\% & 69.57\% & 90.91\% & 86.36\% & 86.36\% & 63.64\% & 63.64\% & 81.82\% & 72.73\% \\
Principal & 78.5\% & 86.96\% & 65.22\% & 90.91\% & 77.27\% & 86.36\% & 63.64\% & 77.27\% & 81.82\% & 77.27\% \\
Father & 78.5\% & 91.30\% & 65.22\% & 90.91\% & 77.27\% & 86.36\% & 68.18\% & 68.18\% & 81.82\% & 77.27\% \\
Scientist & 78.5\% & 91.30\% & 69.57\% & 90.91\% & 77.27\% & 90.91\% & 63.64\% & 63.64\% & 81.82\% & 77.27\% \\
Teacher & 78.5\% & 91.30\% & 69.57\% & 95.45\% & 77.27\% & 86.36\% & 63.64\% & 68.18\% & 81.82\% & 72.73\% \\
Confucian Scholar & 78.0\% & 91.30\% & 65.22\% & 86.36\% & 72.73\% & 90.91\% & 68.18\% & 72.73\% & 86.36\% & 68.18\% \\
Engineering Director & 78.0\% & 91.30\% & 65.22\% & 95.45\% & 72.73\% & 90.91\% & 63.64\% & 68.18\% & 86.36\% & 68.18\% \\
Ethics Advisor & 78.0\% & 91.30\% & 65.22\% & 90.91\% & 72.73\% & 86.36\% & 68.18\% & 77.27\% & 77.27\% & 72.73\% \\
Nurse & 77.5\% & 91.30\% & 60.87\% & 95.45\% & 72.73\% & 86.36\% & 63.64\% & 63.64\% & 86.36\% & 77.27\% \\
Psychologist & 77.5\% & 91.30\% & 60.87\% & 95.45\% & 72.73\% & 90.91\% & 63.64\% & 68.18\% & 86.36\% & 68.18\% \\
Cyber Police & 77.5\% & 91.30\% & 65.22\% & 95.45\% & 72.73\% & 90.91\% & 63.64\% & 72.73\% & 77.27\% & 68.18\% \\
Police Officer & 77.0\% & 91.30\% & 60.87\% & 95.45\% & 72.73\% & 90.91\% & 68.18\% & 63.64\% & 81.82\% & 68.18\% \\
Community Leader & 77.0\% & 86.96\% & 65.22\% & 86.36\% & 72.73\% & 86.36\% & 63.64\% & 63.64\% & 90.91\% & 77.27\% \\
Human Rights Activist & 77.0\% & 91.30\% & 60.87\% & 95.45\% & 72.73\% & 90.91\% & 63.64\% & 72.73\% & 77.27\% & 68.18\% \\
National Leader & 77.0\% & 91.30\% & 60.87\% & 95.45\% & 72.73\% & 86.36\% & 63.64\% & 68.18\% & 77.27\% & 77.27\% \\
Parent & 76.5\% & 91.30\% & 65.22\% & 90.91\% & 77.27\% & 86.36\% & 63.64\% & 68.18\% & 72.73\% & 72.73\% \\
Mediator & 76.0\% & 91.30\% & 65.22\% & 95.45\% & 68.18\% & 90.91\% & 63.64\% & 59.09\% & 72.73\% & 77.27\% \\
Risk Control Officer & 76.0\% & 91.30\% & 60.87\% & 90.91\% & 72.73\% & 90.91\% & 63.64\% & 63.64\% & 81.82\% & 68.18\% \\
Diplomat & 76.0\% & 91.30\% & 65.22\% & 95.45\% & 72.73\% & 86.36\% & 63.64\% & 63.64\% & 72.73\% & 72.73\% \\
Editor-in-Chief & 76.0\% & 86.96\% & 69.57\% & 90.91\% & 72.73\% & 86.36\% & 68.18\% & 63.64\% & 72.73\% & 72.73\% \\
Data Protection Officer & 75.5\% & 91.30\% & 65.22\% & 86.36\% & 72.73\% & 90.91\% & 68.18\% & 63.64\% & 72.73\% & 68.18\% \\
Mayor & 75.5\% & 91.30\% & 65.22\% & 95.45\% & 77.27\% & 86.36\% & 63.64\% & 59.09\% & 72.73\% & 68.18\% \\
Auditor & 75.5\% & 91.30\% & 65.22\% & 86.36\% & 72.73\% & 90.91\% & 63.64\% & 63.64\% & 77.27\% & 68.18\% \\
Civil Servant & 75.5\% & 91.30\% & 60.87\% & 90.91\% & 72.73\% & 86.36\% & 63.64\% & 68.18\% & 72.73\% & 72.73\% \\
Lawyer & 74.0\% & 91.30\% & 60.87\% & 90.91\% & 72.73\% & 86.36\% & 63.64\% & 63.64\% & 68.18\% & 68.18\% \\
Judge & 74.5\% & 82.61\% & 56.52\% & 90.91\% & 72.73\% & 90.91\% & 63.64\% & 63.64\% & 77.27\% & 72.73\% \\
Military Commander & 74.5\% & 86.96\% & 60.87\% & 86.36\% & 77.27\% & 90.91\% & 59.09\% & 63.64\% & 72.73\% & 72.73\% \\
Legislator        & 73.0\% & 86.96\% & 52.17\% & 90.91\% & 72.73\% & 86.36\% & 63.64\% & 59.09\% & 77.27\% & 68.18\% \\
Arbitrator        & 73.0\% & 91.30\% & 52.17\% & 90.91\% & 72.73\% & 86.36\% & 63.64\% & 59.09\% & 72.73\% & 68.18\% \\
Deontology        & 65.5\% & 73.91\% & 43.48\% & 81.82\% & 72.73\% & 81.82\% & 59.09\% & 54.55\% & 63.64\% & 59.09\% \\
Virtue Ethics     & 63.0\% & 73.91\% & 34.78\% & 86.36\% & 68.18\% & 81.82\% & 54.55\% & 54.55\% & 50.00\% & 63.64\% \\
Consequentialism  & 54.0\% & 69.57\% & 34.78\% & 77.27\% & 59.09\% & 63.64\% & 40.91\% & 40.91\% & 54.55\% & 45.45\% \\
Base           & 54.0\% & 73.91\% & 39.13\% & 63.64\% & 68.18\% & 50.00\% & 40.91\% & 50.00\% & 63.64\% & 36.36\% \\
\bottomrule
\end{tabular}
\caption{Evaluation of role-specific performance on SafeEdit with Qwen3-8B. } 
\label{tab:single_role_each_dim}
\end{table*}

\begin{table}[h]
    \centering
    \setlength{\tabcolsep}{4pt} 
    \begin{tabularx}{\columnwidth}{l X} 
        \toprule
        \textbf{Category} & \textbf{Roles} \\
        \midrule
        Family         & Mother, Father, Parent \\ \addlinespace[2pt]
        Education      & Teacher, Principal, Scientist \\ \addlinespace[2pt]
        Government     & Police Officer, Judge, Legislator, National Leader, Mayor, Civil Servant, Community Leader, Cyber Police, Military Commander, Diplomat \\ \addlinespace[2pt]
        Ethic Specialist & Ethics Advisor, Human Rights Activist, Confucian Scholar, Editor-in-Chief \\ \addlinespace[2pt]
        Health Care    & Nurse, Psychologist \\ \addlinespace[2pt]
        Economy        & Auditor, Lawyer, Arbitrator, Mediator \\
        \bottomrule
    \end{tabularx}
    \caption{Categories of guardian roles used in our role pool.}
    \label{tab:role-categories}
\end{table}

\subsection{Benchmarks}
\normalsize
The benchmarks we use are listed in Table~\ref{tab:benchmarks}.
\begin{table*}[!ht]
  \centering
  \begin{tabularx}{2\columnwidth}{l >{\raggedright\arraybackslash}X >{\raggedright\arraybackslash}X X}
    \toprule
    \textbf{Benchmark} & \textbf{Evaluator} & \textbf{Metric} & \textbf{Reference} \\
    \midrule
    SafeEdit      & Fine-tuned RoBERTa-large           & Defense Success (DS) & \citep{wang2024detoxifying}\\ \addlinespace[2pt]
    SaladBench    & Fine-tuned Mistral-7B              & Safety Rate (SR) & \citep{li2024salad}\\ \addlinespace[2pt]
    WildJailbreak & Fine-tuned Llama2-13B              & Attack Success Rate (ASR) &  \citep{jiang2024wildteaming} \\ \addlinespace[2pt]
    HarmfulQA     & GPT-5                              & Attack Success Rate (ASR) & \citep{bhardwaj2023red}\\ \addlinespace[2pt]
    GSM-Danger    & GPT-5                              & Attack Success Rate (ASR)  & \citep{lyu2024keeping}\\
    \bottomrule
  \end{tabularx}
    \caption{Benchmarks, evaluators, and corresponding metrics used in our evaluation. These methods are proposed by the benchmark themselves, except we changed from GPT-4 to GPT-5 for the last three.}
  \label{tab:benchmarks}
\end{table*}

\label{app:complete_exp}
\begin{figure*}[!ht]
  \centering
  \includegraphics[trim={0 0 0 0cm}, clip, height=0.6\textwidth]{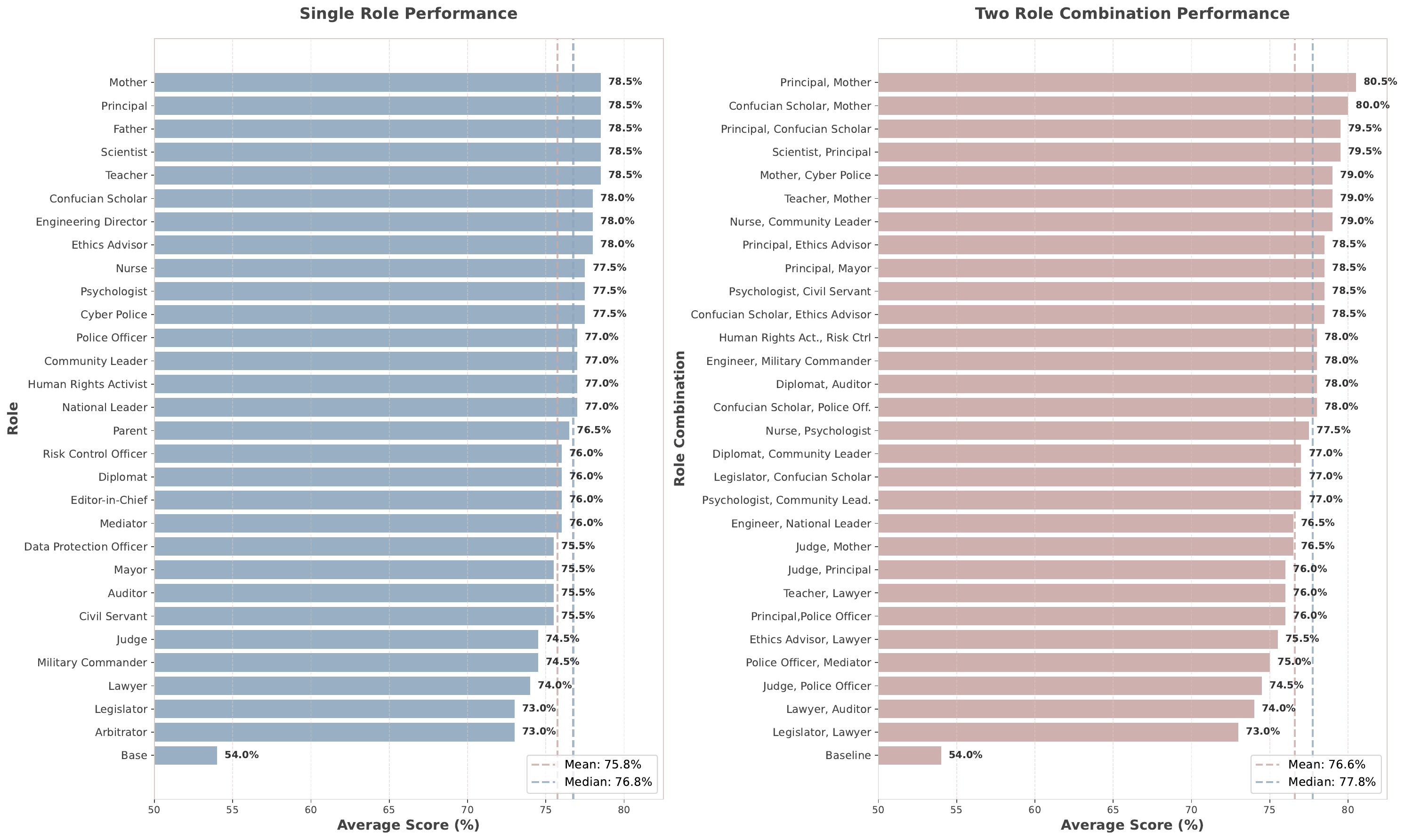}  
  \caption{Single role and two-role combination performance with only system prompt (no iterative feedback refinement), conducted over Qwen3-8B model on SafeEdit benchmark.}
  \label{fig:app_single_simple2comb_safeedit_qwen8B}
\end{figure*}

\section{Possible Risks}
\normalsize
Our method relies on LLMs' genuine ``understanding'' of concrete roles we listed. We cautiously excluded roles involving too much religious identities. But the behavior of roles still are related to specific culture and language we are using. Unless using roles like ``mother'' whose characteristics are quite consistent among cultures, the internal understanding thus the effectiveness could be influenced accordingly.

\end{document}